\documentclass[12pt, a4paper]{article}
\usepackage[utf8]{inputenc}
\usepackage{dcolumn,lscape}%,setspace}
\usepackage{amsmath,longtable,multicol,dcolumn,tabularx,graphicx,amssymb}
\usepackage{exscale,amsthm,multirow,rotating}

\usepackage{bm}
\usepackage{caption}
\usepackage{subcaption}
\usepackage{hyperref}

\usepackage{natbib}
\usepackage{bbm}
\usepackage{tikz,dsfont,float}
\usetikzlibrary{calc}
\usetikzlibrary{positioning}

\begin{document}

\def\spacingset#1{\renewcommand{\baselinestretch}%
{#1}\small\normalsize} \spacingset{1}

%%%%%%%%%%%%%%%%%%%%%%%%%%%%%%%%%%%%%%%%%%%%%%%%%%%%%%%%%%%%%%%%%%%%%%%%%%%%%%

\title{\bf Spatiotemporal modelling of PM$_{2.5}$ concentrations in Lombardy (Italy) -- A comparative study}
  \author{Philipp Otto\\
  \textit{University of Glasgow} 
  \and 
  Alessandro Fusta Moro\\
  \textit{University of Bergamo}
  \and 
  Jacopo Rodeschini\\
  \textit{University of Bergamo}
  \and 
  Qendrim Shaboviq\\
  \textit{Leibniz University Hannover}
  \and 
  Rosaria Ignaccolo\\
  \textit{University of Turin}
  \and 
  Natalia Golini\\
  \textit{University of Turin}
  \and 
  Michela Cameletti\\
  \textit{University of Bergamo}
  \and 
  Paolo Maranzano\\
  \textit{University of Milan-Bicocca, Fondazione Eni Enrico Mattei (FEEM)}
  \and 
  Francesco Finazzi\\
  \textit{University of Bergamo}
  \and 
  Alessandro Fassò\\
  \textit{University of Bergamo}}
  \maketitle

\begin{abstract}
	This study presents a comparative analysis of three predictive models with an increasing degree of flexibility: hidden dynamic geostatistical models (HDGM), generalised additive mixed models (GAMM), and the random forest spatiotemporal kriging models (RFSTK). These models are evaluated for their effectiveness in predicting PM$_{2.5}$ concentrations in Lombardy (North Italy) from 2016 to 2020. Despite differing methodologies, all models demonstrate proficient capture of spatiotemporal patterns within air pollution data with similar out-of-sample performance. Furthermore, the study delves into station-specific analyses, revealing variable model performance contingent on localised conditions. Model interpretation, facilitated by parametric coefficient analysis and partial dependence plots, unveils consistent associations between predictor variables and PM$_{2.5}$ concentrations. Despite nuanced variations in modelling spatiotemporal correlations, all models effectively accounted for the underlying dependence. In summary, this study underscores the efficacy of conventional techniques in modelling correlated spatiotemporal data, concurrently highlighting the complementary potential of Machine Learning and classical statistical approaches.
\end{abstract}

\noindent%
{\it Keywords:} Air pollution, spatiotemporal process, geostatistics, machine learning, hidden dynamic geostatistical model, generalised additive mixed model, random forest spatiotemporal kriging.
% \vfill

% \newpage
\spacingset{1.45} % DON'T change the spacing!

\section{Introduction}\label{sec::introduction}

%1. AgrImOnIA project
The Lombardy region, situated in the heart of the Po Valley in Northern Italy, is known to be highly polluted due to the natural barrier created by the Alps hindering the dispersion of air pollutants \citep[see, e.g.,][]{pernigotti2012impact}. Fine particulate matter (PM\textsubscript{2.5}) has been identified as the most hazardous air pollutant \citep{airEU2022}, representing a mix of air pollutants with a diameter less than 2.5$\mu$m \citep[see also][for a review]{jerrett2005review}. Information about the air quality dynamics is essential for decision-makers to effectively mitigate adverse effects. Statistical and machine learning models can offer valuable insights into this behaviour and its consequences, including the identification of pollution sources and the factors influencing its behaviour and forecasting future pollution levels under various scenarios, such as changes in emissions or weather patterns. Moreover, the combination of different modelling techniques may further enhance the results. By leveraging the strengths of each approach, decision-makers can gain a more comprehensive understanding of air pollution and make more informed choices for mitigation strategies.

This paper compares three statistical and machine learning models with varying degrees of flexibility to elucidate daily PM$_{2.5}$ concentrations in the Lombardy region. More precisely, hidden dynamic geostatistical models (HDGM), generalised additive mixed models (GAMM), and random forest spatiotemporal kriging (RFSTK) were utilised to describe the relationships between a large set of predictors and PM$_{2.5}$ concentrations. All three models employed in this study have been specifically developed to handle spatiotemporal data. HDGM incorporates a latent variable to capture spatiotemporal dependence, while external factors are included in a linear manner within the model. Conversely, GAMM allows for the nonlinear impact of exogenous predictors, which are estimated using splines. It incorporates spatiotemporal dependence by utilising a smoothing spline for spatial variation and a first-order autoregressive process for temporal dependence. Lastly, RFSTK employs a random forest (RF) to model the nonlinear effects of predictors and then a spatiotemporal kriging model to account for the possible spatiotemporal dependence. 

These models are frequently applied in diverse areas. First, HDGM has been primarily used for air pollution studies (see, e.g., \citealt{najafabadi2020spatiotemporal,taghavi2020concurrent} for air pollution in Iran, and \citealt{maranzano2023adaptive,fasso2022spatiotemporal,calculli2015maximum} for Italy). Notably, there are further applications in other fields, such as modelling bike-sharing data or coastal profiles \citep[see][]{piter2022helsinki,otto2021statistical}. HDGM is a linear mixed effects model with a specific structure of the random effects capturing the spatiotemporal dynamics of environmental data, which are widely applied in diverse areas (see, e.g., \citealt{jiang2007linear} for an overview). Second, GAMM has been employed in various areas, such as ecology \citep{knape2016decomposing,kneib2011general}, psychology \citep{bono2021report}, economics \citep[][]{fahrmeir2001bayesian}, psycholinguistics \citep{baayen2017cave}, or event studies \citep{maranzano2023spatiotemporal}. Third, needless to say, models based on decision trees have demonstrated their effectiveness in capturing complex patterns in different fields (see, e.g., \citealt{belgiu2016random} for an overview in remote sensing, or \citealt{qi2012random} for bioinformatics), particularly in combination with kriging approaches, e.g., in environmental \citep{sekulic2020random,chen2019assessment,guo2015digital}, or air pollution studies \citep{liu2018improve,liu2019revisiting}. We refer the interested reader to the systematic literature review of \citep{patelli2023path} for a structured overview of these approaches. Furthermore, an interesting new approach is to use deep neural networks for the prediction and interpolation of spatial data \citep[see][]{NAG2023100773,daw2023reds}.
Hybrid models, integrating different models in one single framework and exhibiting good robustness and adaptability, can combine the advantages of different models during the different stages of the modelling phase. Their adoption is rapidly increasing in various fields and predictions, including PM$_{2.5}$ \citep[e.g.,][]{bai2022novel,tsokov2022hybrid,sun2022hybrid,wang2019digital,ding2021hybrid}, greenhouse gas emissions \citep{javanmard2022hybrid}, tea yield \citep{jui2022spatiotemporal}, depopulation in rural areas \citep{jato2023statistical}, or disease monitoring \citep{kishi2023characteristic} and calibration of citizen-science air quality data \citep{bonas2021calibration}.

All three models account for the intrinsic spatial, temporal, and spatiotemporal dynamics of the PM$_{2.5}$ concentrations. This temporal and spatial dependence arises from the persistence of the particles in the atmosphere over a certain time and, simultaneously, from the displacement and spread of the particles to nearby areas, e.g., by wind \citep{merk2020estimation}. Previous studies successfully employed several statistical models to model air pollution scenarios in Northern Italy, such as generalised additive models \citep{bertaccini2012modeling}, Bayesian hierarchical modes based on the stochastic partial differential equation approach \citep{cameletti2013spatio, fioravanti2021spatio}, or random forests \citep{stafoggia2019estimation}. In a comparative study for Northern Italy, \cite{cameletti2011comparing} studied the effectiveness of different statistical models in a Bayesian framework. Machine learning algorithms, including random forests, are adept at capturing nonlinearities and interactions. Still, when applied to air quality modelling, the spatiotemporal nature of the phenomenon is often ignored \citep[see, e.g.,][]{fox2020comparing}. Consequently, the model's performance deteriorates, with worse outcomes than those obtained from Kriging with External Drift (KED), considered standard for modelling spatiotemporal phenomena. KED shows better results than random forest in Lombardy \citep{fusta2022ammonia} and in the USA \citep{berrocal2020comparison}. On the other hand, machine learning algorithms outperform classical models if spatiotemporal dependence is not considered at all \citep{kulkarni2022model}. \cite{lu2023comparison} compared geostatistical and ML models for NO$_{2}$ concentrations in Germany. Despite the limited number of studies comparing geostatistical and ML models, this subject is gaining increasing interest because the comparison provides valuable insights into the dynamics of the process, as we will illustrate below.

The remaining sections of this paper are organised as follows. Section \ref{sec::data} describes the general framework of the study and the data set used for our comparisons. Then, we explain the theoretical background of all considered models in Section \ref{sec:methods}. The comparative study is presented in Section \ref{sec::results}, including fitting procedure (Section \ref{sec::results_fitting}), residual analysis (Section \ref{sec::results_residuals}), prediction performances within the cross-validation scheme (Section \ref{sec::results_cv_comparison}), and model interpretation (Section \ref{sec::results_interpretation}). Section \ref{sec::conclusion} concludes the paper.

\section{Data}\label{sec::data}

Our comparative analysis is based on the \textit{Agrimonia data set}, a comprehensive daily spatiotemporal data set for air quality modelling available open-access on Zenodo \citep{fassodataset}. Specifically, it includes air pollutant concentrations and important covariates for all 141 stations of the air quality monitoring network in the Lombardy region and a 30 km buffer zone around the administrative boundaries. The data originates from multiple sources with different temporal and spatial resolutions. Using suitable aggregation and interpolation techniques described in \cite{fasso2022agrimonia}, the \textit{Agrimonia data set} is available on a daily basis for all ground-level measurement stations in the study area. It spans six years, from 2016 to 2021, and includes daily air pollutant concentrations, weather conditions, emissions flows, land use characteristics, and livestock densities. We summarise all variables considered in this study in Table \ref{tab::sub_variables}, including their main descriptive statistics. 

The response variable is the $\text{PM}_{2.5}$ concentration at the ground described in ensuing Section \ref{sec::PM25} in more detail, while the selected remaining variables of the \textit{Agrimonia data set} serve as explanatory variables or features. They are summarised and motivated in Section \ref{sec::re}.

\subsection{$\text{PM}_{2.5}$ concentrations}\label{sec::PM25}

\begin{figure}
    \centering
    \includegraphics[width=1\textwidth]{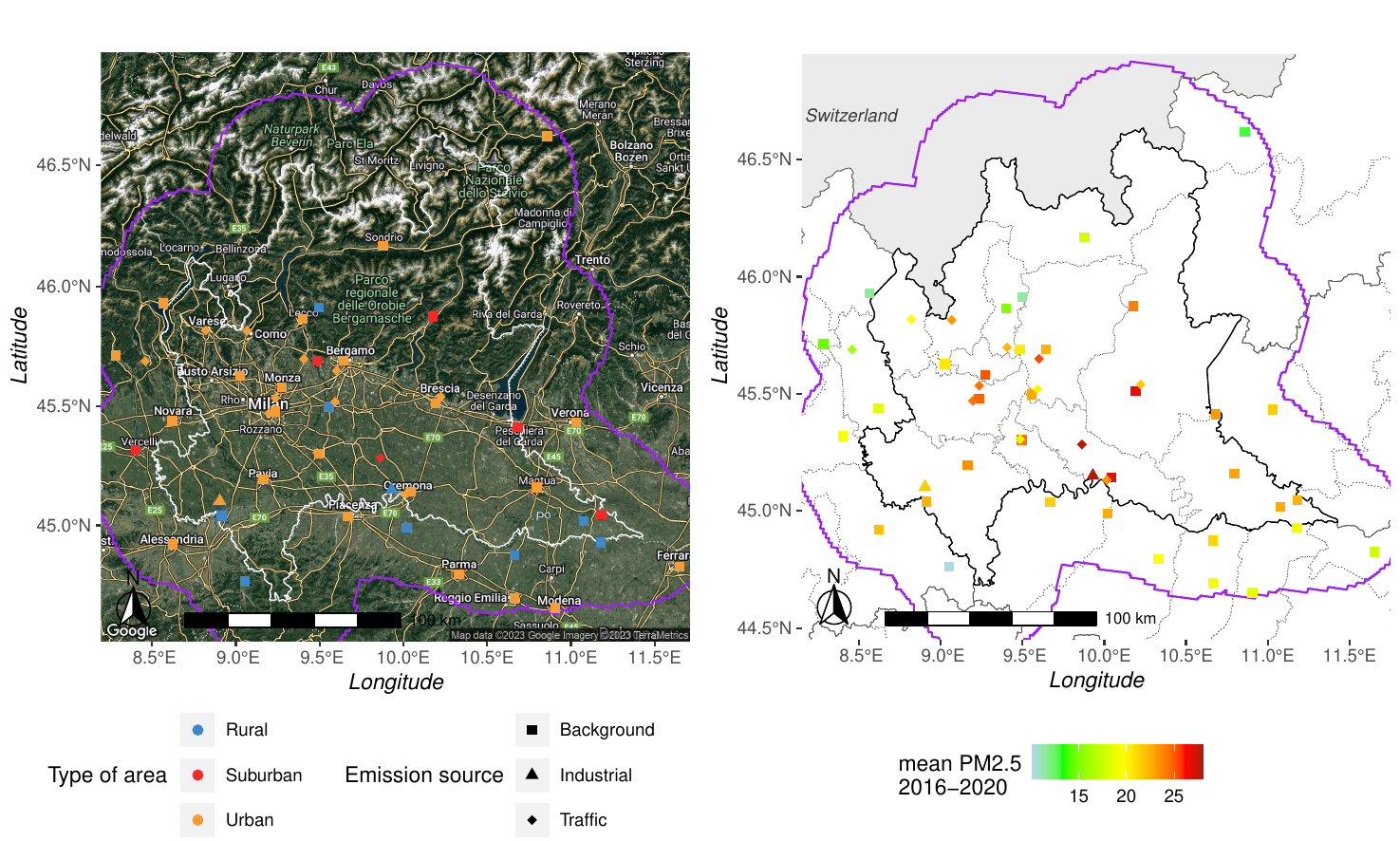}
    \caption{Map of the 49 $\text{PM}_{2.5}$ monitoring stations extracted from the \textit{Agrimonia data set}. The purple line represents a 0.3$^\circ$ buffer around the administrative boundaries of the Lombardy region, the latter represented by the white line. Left: Stations are coloured according to the type of area. The shape indicates the main emission sources. Right: the stations are coloured according to the 2016-2020 average PM$_{2.5}$ concentrations [$ \mu g m^-3$] over 2016-2020.}
    \label{fig:stations}
\end{figure}

The \textit{Agrimonia data set} includes daily observations of several atmospheric pollutants retrieved from the Italian air quality monitoring network. Not all monitoring stations are equipped with the same sensors, so we have excluded locations where stations were not measuring PM\textsubscript{2.5}, which is the target pollutant of this study. The 49 remaining stations are depicted in Fig. \ref{fig:stations} (left) along with information about the type of surrounding area (rural, suburban, urban) and the primary nearest emission source (background, industrial, traffic), according to the EU classification \citep{EUclass}. To depict the spatial variation of the PM concentrations, we coloured the stations according to the average daily concentration across the entire time period on the right-hand map in Fig. \ref{fig:stations}. 

\begin{figure}
    \centering
    \includegraphics[width=1\textwidth]{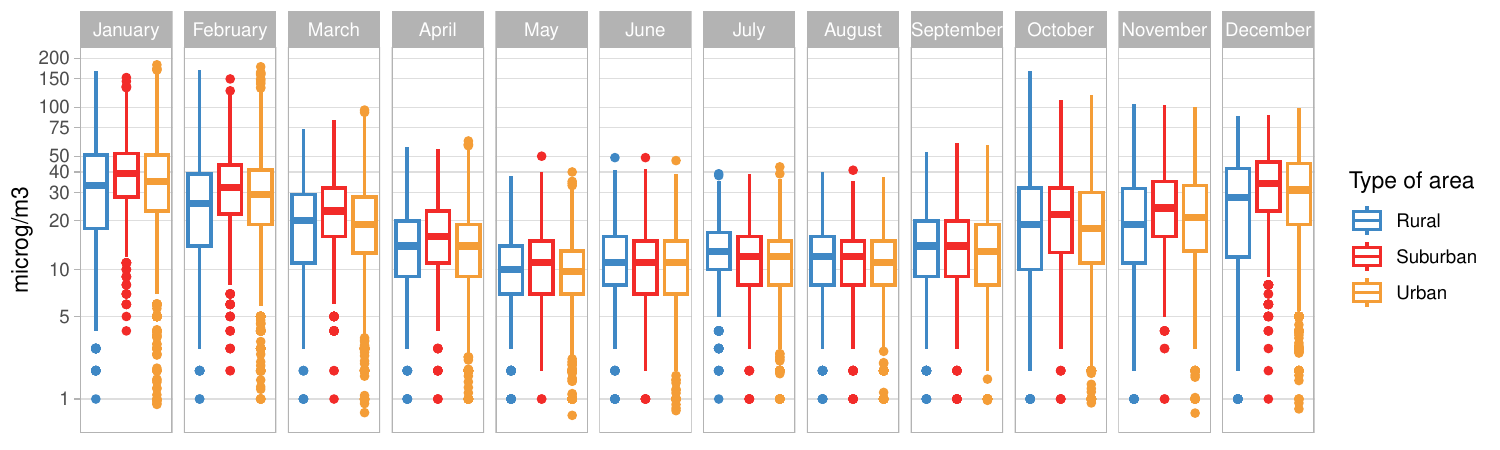}
    \caption{Monthly boxplots of PM\textsubscript{2.5} concentrations (on a log scale) measured in Lombardy, including the buffer area.}
    \label{fig::pm25boxplot}
\end{figure}

We consider the period from 2016 to 2020. The temporal variation of the observed PM\textsubscript{2.5} concentrations, grouped by months and by type of area, is displayed through a series of boxplots in Fig. \ref{fig::pm25boxplot}. The colours are chosen according to the type of the surrounding area. Not surprisingly, there is a clear seasonality with higher concentrations in winter due to meteorological conditions resulting in reduced air circulation. The median concentrations range between 10$\mu g m^{-3}$ and 40$\mu g m^{-3}$ across the year. Thus, throughout the year, the median concentration was beyond the threshold of 5$\mu g m^{-3}$ considered hazardous by the World Health Organisation guidelines \citep{WHOlimit}. From Fig. \ref{fig::pm25boxplot}, it is clear that all different types of areas are similarly affected by poor air quality. This spatial homogeneity is also visually confirmed in the map of Fig. \ref{fig:stations} where clusters of similar neighbouring concentrations can be seen, suggesting a pronounced spatial dependence.

\begin{figure}
    \centering
    \includegraphics[width=.5\textwidth]{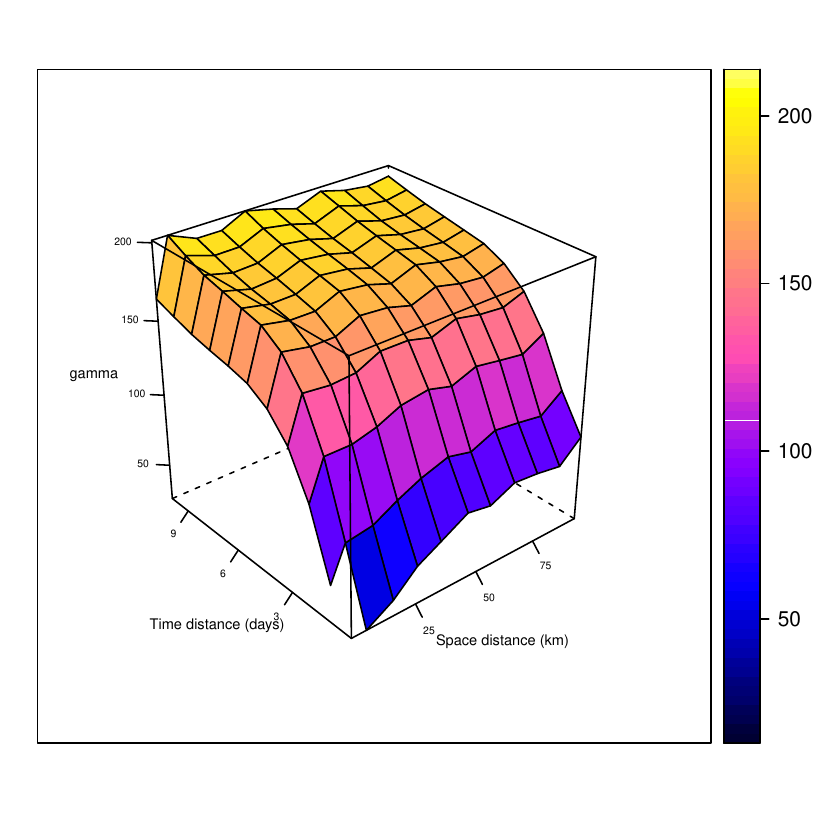}
    \caption{Spatiotemporal variogram of the PM\textsubscript{2.5} concentrations.}
    \label{fig::var}
\end{figure}

To explore a possible spatiotemporal correlation, we estimate a spatiotemporal variogram $\gamma(h,\tau)$ based on the sample variance of observations within certain distance ranges in space and time $h$ and $\tau$ \citep[see, e.g.,][]{cressie2015statistics}. Smaller values of the variogram for smaller distances indicate (short-term) statistical dependence. The spatiotemporal variogram of observed PM2.5 concentrations is depicted in Fig. \ref{fig::var}. As expected, the variogram identifies an apparent correlation of the PM\textsubscript{2.5} concentrations across time and space. More precisely, the values of the variogram for the first temporal lags indicate a pronounced temporal dependence within the first 5-6 days, i.e., approximately one week. Furthermore, we observe a noticeable spatial dependence since the variogram increases with increasing spatial distances. It is important to note that this variation still includes spatial and temporal seasonalities and variations caused by exogenous factors.

\subsection{Regressors}\label{sec::re}

All models, which will later be used for the comparison, share the same set of regressors, including weather conditions and livestock densities. Based on an extensive literature review on air pollution modelling, we carefully selected key weather variables and incorporated information regarding local-scale animal breeding. The specific variables considered are presented in Table \ref{tab::sub_variables} and their corresponding descriptive statistics. Furthermore, monthly indicator variables are included to capture the seasonality, as highlighted by Fig. \ref{fig::pm25boxplot}. Below, we will provide a brief motivation for each variable and the descriptive statistics in Table \ref{tab::sub_variables} to offer an intuitive understanding of the explanatory variables.

\begin{table}
    \caption{Variables selected from the \textit{Agrimonia data set}. Names are consistent with the data set \citep{fassodataset} and the accompanying data descriptor \citep{fasso2022agrimonia}.}
  \label{tab::sub_variables} 
  \footnotesize
  \centering
\begin{tabular}{@{\extracolsep{1pt}}p{8cm} rrrr} 
\hline
\textbf{Variable} - Description [unit of measure] & \multicolumn{1}{c}{Min} & \multicolumn{1}{c}{Mean} & \multicolumn{1}{c}{St. Dev.} & \multicolumn{1}{c}{Max} \\ 
\hline
\textbf{Altitude} \linebreak Height in relation to sea level [$m$] & $4$ & $171.458$ & $200.004$ & $1,194$ \\[.4cm]
\textbf{AQ\_pm25} \linebreak Fine particulate matter concentrations [$\mu g m^{-3}$] & $0.506$ & $20.896$ & $16.353$ & $182$ \\[.4cm]
\textbf{WE\_temp\_2m} \linebreak Air temperature at 2 m [°C]& $-$$20.650$ & $12.916$ & $8.233$ & $32.880$ \\[.4cm]
\textbf{WE\_tot\_precipitation} \linebreak Total precipitation [$m$] & $0$ & $0.003$ & $0.008$ & $0.172$ \\[.4cm]
\textbf{WE\_rh\_mean} \linebreak Relative humidity [\%] & $19.490$ & $74.433$ & $12.299$ & $99.520$ \\[.4cm]
\textbf{WE\_wind\_speed\_100m\_mean} \linebreak Average wind speed at 100m [$m/s$] & $0.564$ & $2.550$ & $1.326$ & $11.930$ \\[.4cm]
\textbf{WE\_blh\_layer\_max} \linebreak Daily maximum height of the air mixing layer layer [$m$] &  $13.790$ & $1,039.402$ & $556.877$ & $4,421.000$ \\[.4cm]
\textbf{LI\_pigs\_v2} \linebreak Average density of pigs bred for the area (10km$^2$) surrounding the measurement stations [$number/km^2$] & $0.022$ & $115.215$ & $159.666$ & $652.100$ \\[.4cm]
\textbf{LI\_bovine\_v2} \linebreak Average density of bovine bred for the area (10km$^2$) surrounding the measurement stations [$number/km^2$] & $1.543$ & $46.241$ & $47.463$ & $178.800$ \\[.4cm]
\textbf{LA\_hvi} \linebreak High vegetation abundance [$ m^2/m^2 $] & $0.861$ & $2.324$ & $0.804$ & $5.034$ \\[.4cm]
\textbf{LA\_lvi} \linebreak Low vegetation abundance [$ m^2/m^2 $] & $0.865$ & $2.208$ & $0.560$ & $3.662$ \\
\hline 
\end{tabular} 
\end{table}

Firstly, several studies have found that weather is a crucial factor in air quality modelling \citep{bertaccini2012modeling,ignaccolo2014kriging,merk2020estimation,fasso2022spatiotemporal,grange2023meteorologically,chang2022spatio}. Changes in weather conditions such as temperature, precipitation, and wind speed and direction can affect atmospheric stability and turbulence, which can influence the transport and deposition of pollutants. For instance, temperature and boundary layer height are usually negatively related to air pollutant concentrations. Similarly, we typically observe reduced PM concentrations during periods with increased precipitation or wind speed. On the contrary, the direction and size of the effect of the relative humidity are still debated, but it undoubtedly affects the PM$_{2.5}$ concentrations \citep{zhang2017impact}.

Secondly, we considered agricultural influences, which appear to impact air quality \citep[e.g.,][]{Thunis2021,lovarelli2020describing}. The Livestock (LI) data used in this study provide information on the average density of pigs and cattle per municipality (expressed as animals per km$^2$) in the vicinity of each station (within a radius of 10km$^2$). Including livestock data is essential to capture the impact of ammonia (NH\textsubscript{3}) emissions on air quality, as livestock farming is the major source of NH\textsubscript{3} emissions (up to 95\%). Therefore, including LI data in air quality modelling can help better understand and mitigate livestock's impact on air pollution levels.

% Finally, monthly indicator variables are included to capture the seasonality, as highlighted by Fig. \ref{fig::pm25boxplot}. All considered variables and their most important descriptive statistics are summarised in Table \ref{tab::sub_variables}. All names are consistent with the \textit{Agrimonia data set}, so the interested reader can refer to \cite{fasso2022agrimonia} for a more detailed description of all variables.

\section{Spatiotemporal statistical models and machine learning techniques}\label{sec:methods}

We consider the PM\textsubscript{2.5} concentrations as realisations of a spatiotemporal stochastic process \{$Z(s,t) : s\in D, t = 1, 2, ..., T$\}, where $D$ is the spatial domain that contains a set of locations $\{s_i : i = 1,..., n\}$ (i.e., the ground-level measurement stations) and the temporal domain is discrete $t = 1,..., T$ (i.e., daily observations). Furthermore, we posit that $Z(s,t)$ might be influenced by external variables related to weather conditions, emissions, or agricultural activities. Throughout the remainder of the paper, the terms regressors, covariates, and features are used interchangeably. The spatiotemporal proximity of observations typically induces statistical dependence, and thus, the selected model candidates should appropriately incorporate this inherent spatiotemporal dependence. To structure the model alternatives, we can decompose the models into three terms, i.e.,
\begin{equation}
   Z(s,t) = S(s,t) + U(s,t) + \varepsilon(s,t),
    \label{eq::stprocess}
\end{equation}
where $S(s,t)$ is the large-scale component including the regressors, $U(s,t)$ includes small-scale spatiotemporal effects, and $\varepsilon(s,t)$ comprises the measurement and modelling errors, which are assumed to be a zero-mean white noise process. 

\subsection{Hidden dynamic geostatistical model}

The first model selected is the HDGM, which serves as the comparative analysis' starting point or baseline method. It is a widely applied geostatistical model, first considered by \cite{huang1996spatio} as an extension of classical mixed-effects models for univariate spatiotemporal data. \cite{calculli2015maximum} extended the HDGM to multivariate data. This modelling approach proved particularly useful for air quality modelling \citep[e.g.,][]{fasso2011maximum,finazzi2013model}, as the comparative study of \citet{cameletti2011comparing} confirmed. The HDGM specifies the large-scale effects as a linear regression model, i.e.,
\begin{equation}
    S(s,t) = \mathbf{X}_\beta(s,t)'\boldsymbol{\beta},\\
    \label{eq::hdgm_fixed_effect}
\end{equation}
where $\boldsymbol{\beta} = (\beta_0, \ldots, \beta_p)'$ is a vector of $p$ fixed-effect coefficients, including the model intercept $\beta_0$, and $\mathbf{X}_\beta(s,t)$ is the $(s, t)$-th entry of the fixed design matrix of the selected covariates/features. In other words, $\mathbf{X}_\beta(s,t)$ is the vector of the observed covariates at location $s$ and time point $t$. 

The spatiotemporal dependence is modelled as small-scale effects by a geostatistical process
\begin{equation}
    U(s, t) = v \xi(s,t) \, ,
\end{equation}
where $v$ is an unknown, homoscedastic scaling factor, which has to be estimated and describes the degree of the small-scale effects. Further, $\xi(s,t)$ is a latent random variable with Markovian temporal dynamics given by
\begin{equation}
\label{eq:hdgm}
    \xi(s,t) = g_{HDGM} \xi(s,t-1) + \eta(s,t), \quad \eta(s,t) \sim GP \, ,
\end{equation}
where $g_{HDGM} \xi(s,t-1)$ is a hidden first-order autoregressive process process with coefficient $g_{HDGM}$. The temporal dependence is separated from the spatial interactions, which are modelled in $\eta(s,t)$. It is worth noting that this implies a separable space-time covariance.
Specifically, $\eta(s,t)$ is a Gaussian process (GP) with zero mean, unit variance, and covariance matrix determined by an exponential spatial correlation function 
\begin{equation}
\label{eq:hdgm_gamma}
    \rho( \lvert \lvert s - s' \rvert \rvert; \theta_{HDGM}) = exp(- \lvert \lvert s - s' \rvert \rvert / \theta_{HDGM})
\end{equation}
with $\theta_{HDGM}$ being the range parameter, $s$ and $s'$ are two distinct spatial locations, and the distance between them is given by the vector norm $\lvert\lvert \cdot \lvert\lvert$. For this study, we will always employ the distance on the great circle, i.e., the length of the geodesic between $s$ and $s'$. The parameters of the random effects process $\xi(s,t)$ are assumed to be in a space leading to a weakly stationary spatiotemporal process. Finally, $\varepsilon(s,t)$ is an identically distributed random error independent across space and time with zero mean and a constant variance $\sigma^2_\epsilon$.

The model parameter set $\bm{\Phi} =\{ \bm{\beta}, g_{HDGM}, \theta_{HDGM}, v, \sigma^2_\epsilon \}$ is estimated by the maximum-likelihood method using an expectation-maximisation (EM) algorithm \citep{calculli2015maximum}. The estimation procedure is computationally implemented in the MATLAB software package D-STEM \citep[see][]{wang2021d}.

\subsection{Generalised additive mixed model}

Compared to generalised additive models (GAM, \citealt{hastie1987generalized}), generalised additive mixed models (GAMM) include a random-effects component to describe correlated response variables, such as time series, spatial or spatiotemporal data. It extends the HDGM by allowing for linear and nonlinear regressive effects in a GAM fashion, i.e., the response variable linearly depends on smooth functions of the predictors. To be precise, the large-scale components are given by
\begin{equation}
   S(s,t) = \mathbf{X}_{linear}(s,t)'\boldsymbol{\beta}_{linear} + \underbrace{\sum_{j=1}^{m}\alpha_{(j)}(\mathbf{X}_{nonlinear,j}(s,t))}_\text{nonlinear effects} % + \sum_{i = 1}^n b_i C(s, s_i)
    \label{eq::fixedgamm}
\end{equation}
with $\mathbf{X}_{linear}(s,t)\boldsymbol{\beta}_{linear}$ being a linear parametric regression term of the first $k$ covariates with a parameter vector $\boldsymbol{\beta}_{linear} = (\beta_0, \beta_1,...,\beta_k)'$, including the intercept term $\beta_0$. Moreover, $\sum_{j=1}^{m}\alpha_{(j)}(\mathbf{X}_{nonlinear,j}(s,t)))$ is an additive term with nonlinear influence functions $\alpha_{(j)} : \mathbb R \rightarrow \mathbb R$ of the $j$-th column in $\mathbf{X}_{nonlinear,j}$ for the remaining $m$ regressors. These nonlinear influences can be estimated along with the other model coefficients, e.g., as regression splines or penalised splines \citep[][]{fahrmeir2004penalized}.

The small-scale effects of the GAMM are specified as a first-order autoregressive model for the temporal dependence and a smooth spatial surface for the spatial dependence, that is, 
\begin{eqnarray}
    U(s,t)   & = & g_{GAMM} (Z(s, t-1) - S(s, t-1) -  C(s)) \, ,  %  \, \qquad \text{with}   \\
    \label{eq::randomgamm2}
\end{eqnarray}
where $g_{GAMM}$ is the parameter representing the temporal dependence, where zero indicates no temporal correlation. The spatial dependence is modelled as a smooth surface $C(s)$, which follows a Gaussian process with an exponential covariance function with the range parameter $\theta_{GAMM}$ \citep[][]{handcock1994approach}. This structure is identical to the spatial term of the random effects model in HDGM as given by equation \eqref{eq:hdgm_gamma}. In our case, $\theta_{GAMM}$ is estimated as proposed by \cite{kammann2003geoadditive}. The model estimation is computationally implemented in the package \texttt{mgcv} available in R \citep{wood2006generalized}.

\subsection{Random forest spatiotemporal kriging}

For the third approach, RFSTK, we increase the flexibility of the model in the large-scale component by considering a random forest (RF) algorithm. In other words, the third hybrid model combines an RF for the large-scale component $S(s,t)$ with that of a spatiotemporal kriging model for $U(s,t)$. The idea traces back to the combination of random forests and kriging, the so-called random forest residual kriging, which -- even if only considering spatial dependence -- showed promising results compared to RF alone \citep[e.g.][]{wang2019digital,viscarra2014mapping}. For spatiotemporal data, RFSTK has been considered to model air quality -- again showing good performances \citep[see][]{zhan2018satellite,shao2020estimating}. 

Random forests are widely used tools in machine learning as an ensemble of multiple decision trees \citep{breiman2001random}. In a regression problem, the prediction of the large-scale effect is obtained by averaging across the predictions of $n_{\text{tree}}$ decision trees, which are trained/estimated from independent bootstrap samples $Z_j^*$ of the input data ($j = 1, \ldots, n_{\text{tree}}$), i.e.,
\begin{equation}
\label{eq::RF}
    S(s_i,t) = \frac{1}{n_{\text{tree}}} \sum_{j = 1}^{n_{\text{tree}}} \widehat{E}(Z_j^*(s_i, t) \mid \mathbf{X}(s_i, t) : i = 1, \ldots, n, t = 1, \ldots, T), 
\end{equation}
where $\widehat{E}(Z_j^*(s_i, t) \mid \cdot)$ is the prediction of the $j$-th decision tree. For regression trees, it is important to notice that the averaging should be done for each region of interest in the covariate space.
% The bootstrap samples are obtained from independent draws and are equally weighted (bagging).

The small-scale model of RFSTK is assumed to be a zero-mean, weakly stationary spatiotemporal Gaussian process 
\begin{equation}
    U(s,t) = \Tilde{\eta}(s,t) \sim GP ,
\end{equation}
where the covariance matrix is obtained from a separable space-time correlation function given by
\begin{equation}\label{sepcov}
    \rho(\lvert \lvert s - s' \rvert \rvert, \lvert t - t'  \rvert)=\rho(\lvert \lvert s - s' \rvert \rvert; \theta_{RFSTK_s}) \cdot \rho(\lvert  t - t'  \rvert; \theta_{RFSTK_t})
\end{equation}
with $\lvert \lvert s - s' \rvert \rvert $ and $\lvert t - t' \rvert $ representing spatial and temporal distances between $(s,t)$ and $(s', t')$, respectively. That is, the exponential correlation functions are equivalent to the spatial correlation function of the HDGM and GAMM, but the other two approaches consider an autoregressive temporal dependence, while the RFSTK employs a continuous correlation function for both the temporal and spatial dependence. The parameters and decision trees are estimated in a two-step procedure. First, predictions for the large-scale component are obtained using RFs, computationally implemented in the R package \texttt{randomForest} \citep{rf_pack}. Second, to adjust the predictions of the RF accounting for space-time interactions, the parameters of the separable space-time correlation function in \eqref{sepcov} are estimated by variography on RF residuals, implemented in the R package \texttt{gstat} \citep{gstat}.

\section{Comparative study} \label{sec::results}

In the subsequent section, we present an application of each methodology on air quality data extracted from the  \textit{Agrimonia data set} (see Section \ref{sec::data}). We start with exploring the model fitting process and examining the residuals. Subsequently, we evaluate the predictive performances of the models through cross-validation. Finally, our attention shifts to the interpretation of each model and its comparison. Drawing upon this comparison, we offer practical suggestions for integrating the modelling outcomes into other environmental analyses.

\subsection{Model fitting}\label{sec::results_fitting}

For the HDGM, we express the large-scale component $S(s,t)$ in the Wilkinson notation \citep{wilkinson} as follows (see Table \ref{tab::sub_variables} for variable acronyms):
%\begin{scriptsize}
%\begin{equation}\begin{split}
%    \text{AQ\_pm25} = & \; \text{Intercept} \\
%    & + \text{Altitude} + \text{Month} \\
% &    & + \text{WE\_wind\_speed\_100m\_mean} + %\text{WE\_temp\_2m} \\
 %   & + \text{WE\_tot\_precipitation}  + \text{WE\_rh\_mean} + \text{WE\_blh\_layer\_max} \\
%    & + \text{LA\_lvi} + \text{LA\_hvi} \\
%    & + \text{LI\_pigs\_v2} + \text{LI\_bovine\_v2}  \, .
%\end{split}\end{equation}
%\end{scriptsize}
%
\begin{equation}\begin{split}
    \text{AQ\_pm25} \sim & \; 1 \\
    & + \text{Altitude} + \text{Month} \\
    & + \text{WE\_wind\_speed\_100m\_mean} + \text{WE\_temp\_2m} \\
    & + \text{WE\_tot\_precipitation}  + \text{WE\_rh\_mean} + \text{WE\_blh\_layer\_max} \\
    & + \text{LA\_lvi} + \text{LA\_hvi} \\
    & + \text{LI\_pigs\_v2} + \text{LI\_bovine\_v2}  \, .
\end{split}\end{equation}
To ensure the comparability of the results across all three model alternatives, we considered all predictors in their original scale. Transformations such as a logarithmic transformation of the response variable did not generally yield better prediction results and model fits. This suggests an additive structure in the large-scale effects, and its coefficients can directly be interpreted as marginal (linear) effects.

The large-scale component $S(s,t)$ of GAMM captures nonlinear effects by estimating a functional relationship between the predictors and the response variable. Identifying predictors requiring nonlinear relationships entailed simulating model residuals from a linear regression model and graphing them alongside their corresponding predictors, including confidence intervals, as suggested in \cite{fasiolo2020scalable}. A smooth nonlinear effect is estimated if a pattern outside the confidence bands is detected. In this study, all continuous variables, except for altitude, required a smooth term represented by a cubic regression spline. The large-scale component $S(s,t)$ of GAMM in \eqref{eq::fixedgamm} is given by
\begin{equation}\begin{split}
    \text{AQ\_pm25} \sim & \; 1 \\
    & + \text{Altitude} + \text{Month} \\
    & + \text{s(WE\_wind\_speed\_100m\_mean)} + \text{s(WE\_temp\_2m)} \\
    & + \text{s(WE\_tot\_precipitation)} + \text{s(WE\_rh\_mean)} + \text{s(WE\_blh\_layer\_max)} \\
    & + \text{s(LA\_lvi)} + \text{s(LA\_hvi)} \\
    & + \text{s(LI\_pigs\_v2)} + \text{s(LI\_bovine\_v2)} \, ,
\end{split}\end{equation}
where $\text{s(}\cdot\text{)}$ denotes a smooth potentially nonlinear function, $\alpha_{(j)}(\cdot)$ in \eqref{eq::fixedgamm}. For our analysis, we have chosen cubic splines.
The spatial dependence is modelled within the GAMM as a two-dimensional smooth surface $C(s)$ governed by a zero-mean Gaussian process with an exponential covariance function, while the temporal dependence is represented by an autoregressive term of order one. The model is estimated through the restricted maximum likelihood method using the package \texttt{mgcv} in R \citep{mgcv}. 

% RFSTK follows a two-step procedure. Initially, predictions for the large-scale component are obtained using RFs, and subsequently, a spatiotemporal kriging model is fitted to capture the remaining spatiotemporal dependence in the RF residuals. 
To optimise the performance of RF, we evaluate the trade-off between prediction accuracy and computation time across various hyperparameter settings. The latter include the number of trees $n_{tree}$, the number of candidate predictors for building each tree, and the size of the final leaves of each tree. Our findings indicate that utilising default settings (500 trees, the number of candidate variables equal to one-third of the number of the predictors, and final leaf size of 5) within the R package \texttt{randomForest} \citep{rf_pack} is suitable.
Then, RF predictions are adjusted by adding RF residual predictions obtained by fitting an ordinary spatiotemporal kriging model by using the R package \texttt{gstat} \citep{gstat}. % The parameters of the separable space-time correlation function in (\ref{sepcov}) are estimated by variography on RF residuals.

To highlight the difference between the model fits, we compared the in-sample (i.e. using the entire data set) predictive performance of all the models, separately for the large-scale component (LS) and the full model (FM), including the space-time effects. It is worth noting that we included $C(s)$ in the constant term of the LS component for GAMM, allowing for direct comparisons with the HDGM fit of the regression terms. The comparison was based on the root mean squared errors (RMSE), mean absolute errors (MAE), and the coefficient of determination $R^2$. %, and computational costs in terms of clock time. 
The results are reported in Table \ref{tab:residuals}. HDGM generally had better prediction capabilities and lower computational costs than the other models when we consider the full model. The RFSTK model had relatively good prediction capabilities but was computationally intensive, particularly when including the spatiotemporal kriging. For our dataset, the GAMM model had the lowest in-sample fit. However, the informativeness of the in-sample results can be questionable due to the potential sensitivity of models to the training data or overfitting. To address this issue, we evaluated prediction performances within a cross-validation scheme, which is explained in Section \ref{sec::results_cv_comparison}. For instance, this analysis revealed that the HDGM could generalise the estimated relation to obtaining outperforming out-of-sample predictions across space, while we observe a serious overfit in the in-sample case due to the flexibility of the random-effects model. We will focus on this result in more detail below.

\begin{table} \centering 
  \caption{In-sample performance of the three models assessed by RMSE (in $\mu g  m^{-3}$), and the adjusted coefficient of determination $R^2$. The fit of the only large-scale component (LS) is compared to the full model (FM).}  % , and the computation time (clock time in $sec$)
  \label{tab:residuals}
\begin{tabular}{lcccccc} 
\hline
& \multicolumn{2}{c}{HDGM} & \multicolumn{2}{c}{GAMM} & \multicolumn{2}{c}{RFSTK} \\
 & LS & FM & LS & FM & LS & FM \\ 
\hline \\[-1.8ex] 
RMSE [$\mu g m^{-3}$] & $11.92$ & $1.814$ & $11.456$ & $8.468$ & $8.245$ & $5.361$ \\ 
$R^2$                 & $0.469$  & $0.988$ & $0.509$  & $0.732$ & $0.746$ & $0.893$ \\ 
% Time [$sec$]          & $0.01$  & $117$   & $2.51$   & $10287$ & $1235$  & $36993$* \\[.1cm]
\hline \\[-1.8ex] 
% \multicolumn{7}{r}{* with 5 parallel processes}
\end{tabular} 
\end{table}

\subsection{Residual analysis}\label{sec::results_residuals}
%distribution (9)

\begin{figure}
    \centering
    \begin{subfigure}[c]{0.9\textwidth}
    \centering
    \includegraphics[width=1\textwidth]{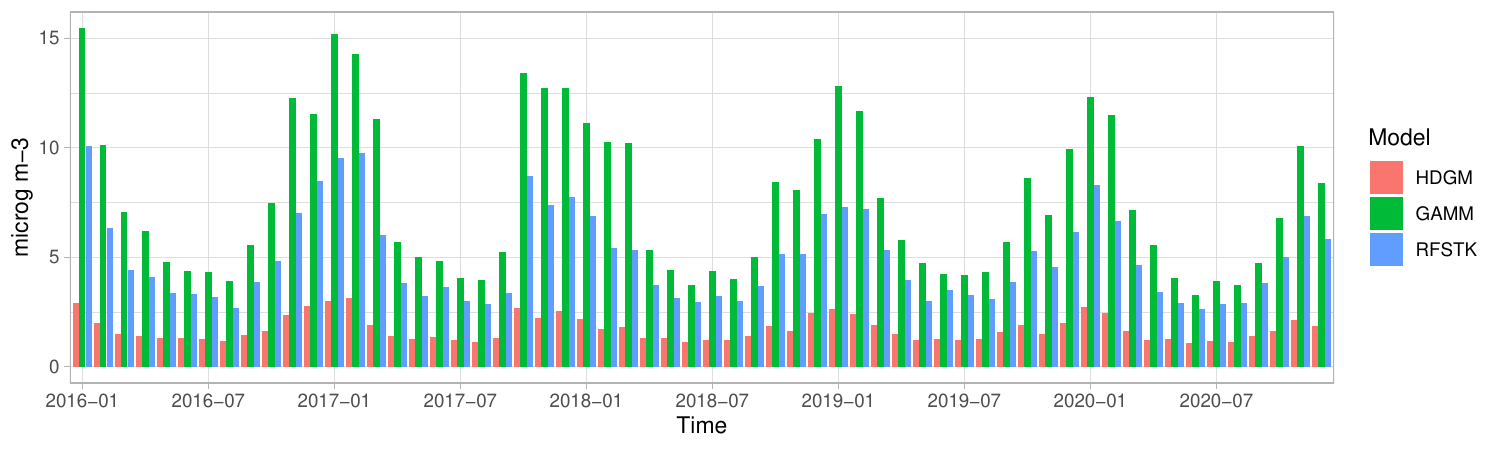}
    \caption{Standard deviation of residuals grouped by months and models.}
    \label{fig:out_res}
    \end{subfigure}
    \begin{subfigure}[c]{0.9\textwidth}
    \includegraphics[width=1\textwidth]{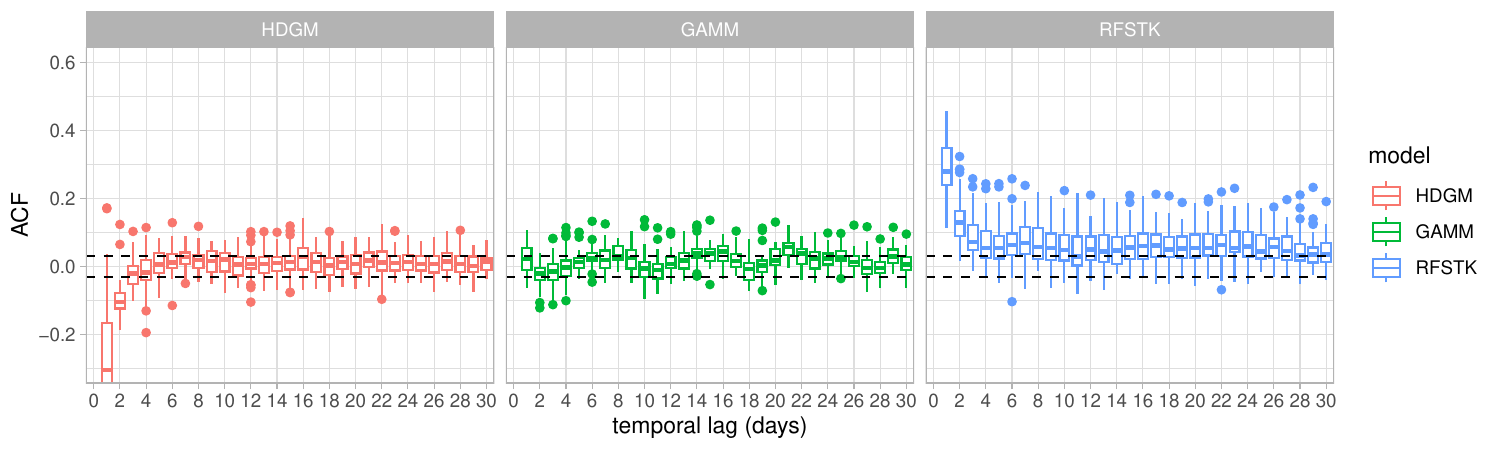}
    \caption{Boxplots (across stations) of residuals' temporal correlograms for the three models.}
    \label{fig:acf_res}
    \end{subfigure}
    \begin{subfigure}[c]{0.9\textwidth}
    \includegraphics[width=1\textwidth]{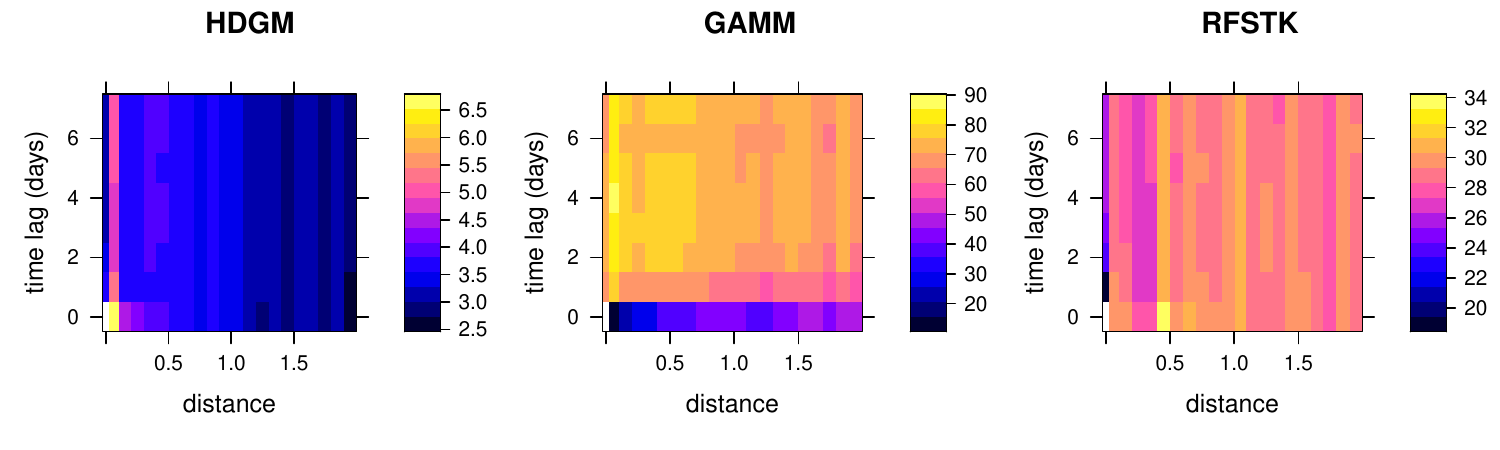}
    \caption{Sample spatiotemporal variogram of the residuals of each model. Spatial distances are expressed in Earth degrees.}
    \label{fig:varST_res}
    \end{subfigure}
    \caption{In-sample residual diagnostics for the three models.}
\end{figure}

The residual distributions are symmetric and slightly leptokurtic, which means there is a greater chance of extreme values than a normal distribution, indicating that all models are less reliable at predicting extreme events. This is not surprising as they are designed to predict the mean level of the distribution and not for modelling the extremes. 
%They are designed to predict the mean or average value of a distribution. Thus, they are not as well-suited to predict extreme events.

The model uncertainties across time, shown in Figure \ref{fig:out_res}, where the standard deviation of the residuals is depicted for each month, reveal that all three models have varying uncertainty throughout the year. More precisely, the PM concentrations in the winter periods could be less accurately predicted than in the summer periods. Therefore, when the models are implemented for forecasting or scenario analysis, it is recommended to use a heteroscedastic model to not underestimate the prediction accuracy in the winter months (or overestimate for the summer period). For instance, spatiotemporal stochastic volatility models could be estimated for the residual process, as demonstrated in \cite{otto2022dynamic} for much simpler mean models. However, in this paper, our focus will be on the comparison of the mean predictions of the three models.

Furthermore, we investigated the spatial and temporal dependence of the residuals estimating temporal autocorrelation functions (ACF) and spatiotemporal variograms. The results are shown in Figures \ref{fig:acf_res} and \ref{fig:varST_res}. Different patterns were observed across the three models. While HDGM shows a small negative correlation at the beginning, indicating a slight overestimation of the temporal dependence, the spatial correlation is satisfactorily captured. On the contrary, GAMM leads to significantly lower temporal correlations in the residuals but does not capture the spatial dependence, as highlighted by the variogram through the bottom line  for time lag 0. The RFSTK is characterised by a more pronounced positive autocorrelation for the first 3-4 lags, as shown by the correlogram (ACF), consistently with the model specification that does not consider an autoregressive term. The spatiotemporal correlation was clearly captured, as confirmed by the flat variogram. It is worth noting that the scales of the variograms differ significantly. This is because the prediction performances of the models also differ significantly in the in-sample case.

\subsection{Cross-validation and comparison of predictive performance}\label{sec::results_cv_comparison}

% Explain LOSO (12)
We employed the leave-one-station-out cross-validation (LOSOCV) scheme, which is a variation of the commonly used leave-one-out cross-validation approach applied in the spatiotemporal framework \cite[e.g.,][]{meyer2018improving,nowak2020improved}. For this method, a complete time series of a single station withheld is not used in the model's training but is used to evaluate the model's prediction performance. In this way, the validation blocks are sufficiently large not to destroy the spatiotemporal dependence. All stations within the Lombardy region were used for validation, except for the station ``Moggio.'' It is located in the mountains with unique climatic conditions that are not well-represented by all other stations. After implementing the LOSOCV approach, we obtained prediction results for the 31 stations included in the validation process. It is worth noting that we applied the identical cross-validation scheme for each model so that the results are directly comparable. 

The prediction performances assessed in the LOSOCV scheme in terms of mean squared errors (MSE), RMSE, MAE, and $R^2$ are summarised in Table \ref{tab:results}. HDGM is confirmed to be the best model, but -- compared to the in-sample residuals in Table \ref{tab:residuals} -- the uncertainty is on a realistic level with an RMSE comparable to the other models. That is, the overfit in the in-sample data did not affect the generalisation ability of the HDGM. This could be due to the linear structure in the large-scale component. While a more flexible model (e.g., random forest or artificial neural networks) could produce extremely bad predictions in areas of insufficient training data or overfitting, the linear structure of the HDGM regression term prevents us from obtaining such extreme predictions. Generally, we observe satisfactory prediction performances for all three model alternatives, with GAMM and the RFSTK approach being in second and third place, respectively. The substantial difference between the in-sample residuals from the model trained on the entire dataset and the errors from the LOSOCV scheme highlights the importance of validating prediction uncertainty through a cross-validation scheme, which accounts for the spatiotemporal nature of the data. Interestingly, the GAMM obtained a similar fit in terms of the coefficient of determination in both the in-sample case and the cross-validation. Thus, we would not overestimate the prediction capabilities when only looking at the in-sample fit. 

\begin{table}
\centering 
  \caption{Prediction performance indices evaluated with the LOSOCV scheme [$\mu g  m^{-3}$]} 
  \label{tab:results} 
\begin{tabular}{l cccc} 
\hline
%& \multicolumn{4}{c}{Test errors} \\
 & MSE & RMSE & MAE & $R^2$ \\ 
\hline
{HDGM} & $35.373$ & $5.948$ & $4.376$ & $0.879$ \\ 
{GAMM} & $78.042$ & $8.834$ & $6.239$ & $0.733$  \\ 
{RFSTK} & $53.099$ & $7.286$ & $5.119$ & $0.819$ \\ 
\hline
\end{tabular} 
\end{table} 

\begin{figure}
    \includegraphics[width=1\textwidth]{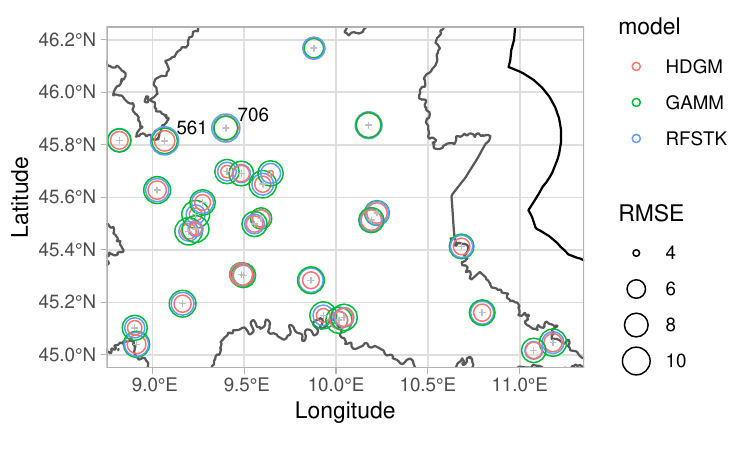}
    \caption{Prediction performances expressed as RMSE calculated for each station within the LOSOCV scheme. The stations ``Lecco - Via Sora'' (ID 706) and ``Como - Via Cattaneo'' (ID 561) are labelled.}
    \label{fig:map_rmse}
\end{figure}

\begin{figure}
    \includegraphics[width=1\textwidth]{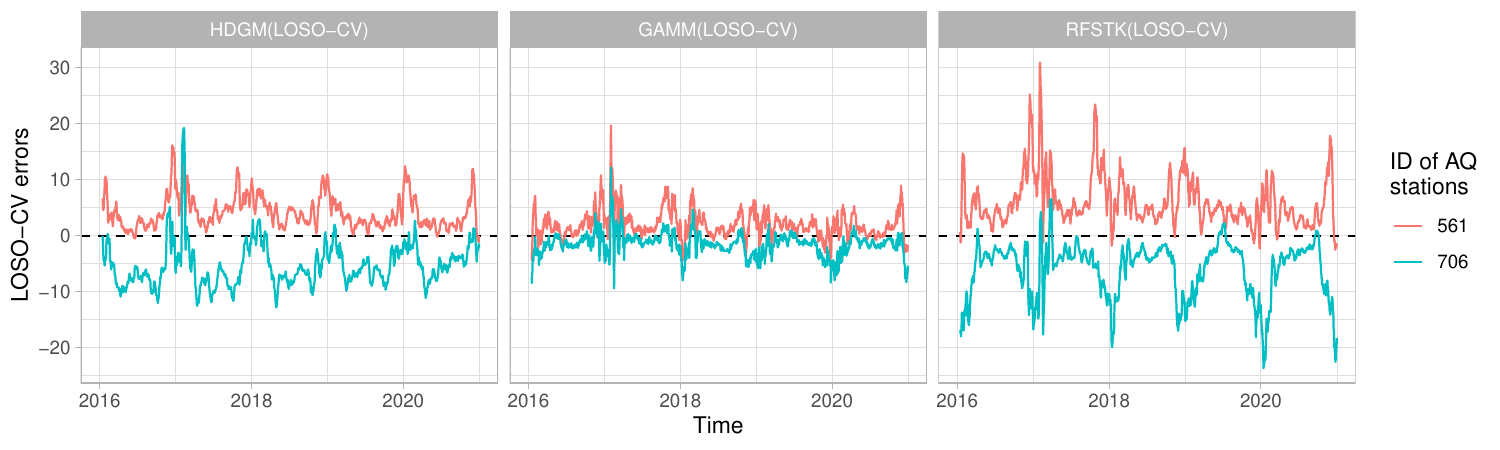}
    \caption{15 days moving averages on the prediction errors (in $\mu g / m^{3}$) for each model (HDGM, GAMM and RFSTK from left to right)}
    \label{fig:err_dist_ts}
\end{figure}

%RMSE by stations (15)
Eventually, we compare the prediction performances for each station separately because we observed that the order of the best-fitting model is not homogeneous across space. For this reason, Figure \ref{fig:map_rmse} displays the cross-validation RMSE on a map by the size of differently coloured circles. That is, the colour of the smallest circle at each station corresponds to the model with the best prediction performance, whereas the largest circles show the worst predictions. Below, we will discuss two selected cases with interesting behaviour: ``Lecco - Via Sora'' (station 706) and ``Como - Via Cattaneo'' (station 561). Furthermore, we depict the cross-validation prediction errors across time for these two stations in Fig. \ref{fig:err_dist_ts}.

HDGM and RFSTK performed worse than GAMM at the station ``Lecco - Via Sora'' (station ID 706). This is because Lecco is characterised by good air quality, but its neighbouring areas are affected by high PM$_{2.5}$ concentrations. The GAMM model, which does not include strong, time-varying spatial interactions, was able to capture this difference in air quality better than the HDGM and RFSTK models. This is confirmed by the fact that the 15-day moving average of the test errors (calculated as observed minus predicted) displayed in Fig. \ref{fig:err_dist_ts} shows that both HDGM and RFSTK overestimate the PM$_{2.5}$ concentrations at this station.

At the other selected station, ``Como - Via Cattaneo'' (station ID 561), the RFSTK model performed the worst while HDGM showed the best performance. The reason may lie in its poor ability to capture temporal dependence well. The concentrations of PM$_{2.5}$ at this station are very stable over time, and the RFSTK model does not fully capture this stability. This is shown by the 15-day moving average of the test errors in Fig. \ref{fig:err_dist_ts}, which shows that RFSTK underestimates PM$_{2.5}$ concentrations, especially in the winter periods, when the air circulation is at its lower limit and temporal stability is the highest.

These results highlight the need to select the model according to the local conditions carefully. The best model for one location may not be the best for another. Moreover, model averaging could additionally improve the predictions.

\subsection{Model interpretation} \label{sec::results_interpretation}

In the following three sections, we delve into the outcomes derived from our models, offering a comprehensive interpretation of each estimated model.

\subsubsection{HDGM}

\begin{table}
\centering \scriptsize
  \caption{Estimated coefficients of the large-scale component of the HDGM model.} 
  \label{tab::hdgm_results}
\begin{tabular}{@{\extracolsep{5pt}} lrrrr} 
\hline 
 Variable & Coefficient & Std. Err. & t-value  & p-value \\ 
\hline
(Intercept) & $39.626$ & $1.399$ & $28.332$ & $<$ 0.0001\\ 
February & -$7.222$ & $1.316$ & $5.490$ & $<$ 0.0001\\ 
March & -$15.298$ & $1.474$ & $10.378$ & $<$ 0.0001\\ 
April & -$21.569$ & $1.552$ & $13.898$ & $<$ 0.0001\\ 
May & -$27.608$ & $1.596$ & $17.295$ & $<$ 0.0001\\ 
June & -$27.662$ & $1.666$ & $16.602$ & $<$ 0.0001\\ 
July & -$27.493$ & $1.690$ & $16.270$ & $<$ 0.0001\\ 
August & -$29.316$ & $1.681$ & $17.436$ & $<$ 0.0001\\ 
September & -$27.611$ & $1.628$ & $16.956$ & $<$ 0.0001\\ 
October & -$20.335$ & $1.555$ & $13.080$ & $<$ 0.0001\\ 
November & -$16.602$ & $1.493$ & $11.117$ & $<$ 0.0001\\ 
December & -$8.526$ & $1.349$ & $6.318$ & $<$ 0.0001\\ 
Altitude & -$0.007$ & $0$ & $20.180$ & $<$ 0.0001\\ 
WE\_wind\_speed\_100m\_mean & -$1.946$ & $0.048$ & $40.520$ & $<$ 0.0001\\ 
WE\_tot\_precipitation & -$159.416$ & $7.145$ & $22.310$ & $<$ 0.0001\\ 
WE\_temp\_2m & $0.505$ & $0.037$ & $13.595$ & $<$ 0.0001\\ 
WE\_rh\_mean & $0.186$ & $0.008$ & $24.188$ & $<$ 0.0001\\ 
WE\_blh\_layer\_max & -$0.003$ & $0$ & $19.628$ & $<$ 0.0001\\ 
LI\_pigs\_v2 & $0.005$ & $0.001$ & $9.288$ & $<$ 0.0001\\ 
LI\_bovine\_v2 & $0.003$ & $0.002$ & $1.147$ & $0.251$ \\ 
LA\_lvi & -$4.441$ & $0.225$ & $19.698$ & $<$ 0.0001\\ 
LA\_hvi & -$0.494$ & $0.141$ & $3.507$ & $<$ 0.0001 \\ 
\hline \\[-1.8ex] 
\end{tabular} 
\end{table}

Table \ref{tab::hdgm_results} summarises the estimated $\bm{\beta}$ parameters of the large-scale component of HDGM. Except for bovine density (LI\_bovine\_v2), all coefficients significantly differ from zero. The signs of the majority of the coefficients are consistent with our expectations: summer months are related to reductions of PM$_{2.5}$ concentrations (about -27/29 $\mu g  m^{-3}$), every $1 m/s$ of wind speed is related to a decrease of 2 $\mu g m^{-3}$ of PM$_{2.5}$, and every $10mm$ of precipitation are related to an expected decrease of $1.6 \mu g m^{-3}$ of PM$_{2.5}$. Moreover, each degree Celsius increase in temperature is related to an increase of $0.5$ $\mu g  m^{-3}$ of PM$_{2.5}$, which seems counter-intuitive at first glance. However, the temperature effect should be interpreted together with the monthly fixed effects. The relative humidity is positively related to PM$_{2.5}$, so high humidity levels ($100\%$) are associated with an increase of 18 $\mu g m^3$ with respect to extremely dry air. The maximum height of the boundary layer is negatively associated with PM$_{2.5}$; every increase of $1000 m$ is related to an expected decrease of 3 $\mu g m^{-3}$ of PM$_{2.5}$.

Regarding the agricultural impact, we observe that the number of pigs in the territory is positively associated, and an increase of 1000 animals per $km^2$ corresponds to an expected increase of 5 $\mu g  m^{-3}$ of PM$_{2.5}$. Both vegetation indices are negatively related to PM$_{2.5}$, while low vegetation (e.g. bushes) has a stronger effect than higher vegetation (e.g. trees).

The small-scale effect of the HDGM is defined by a latent variable $\xi(s,t)$ in \eqref{eq:hdgm}, which has an autoregressive structure of order one and a Gaussian process with exponential covariance function given by \eqref{eq:hdgm_gamma}. The spatial range parameter $\theta_{HDGM}$ describes the decay of the exponential correlation function and is estimated to be equal to $0.79^{\circ}$. Thus, there is a large spatial correlation (i.e., $> 0.37$) for surrounding stations in an area of 80 kilometres. The estimate of the temporal autoregressive parameter is equal to $\hat{g}_{HDGM} = 0.72$. This indicates that the time series has low-frequency components with relatively gradual changes over time. % In other words, the time series does not have a lot of sharp spikes or sudden changes.

\begin{table}
\centering
\caption{Estimated coefficients of the large-scale component of the GAMM. Linear relationships (A) are identified by the $\beta$ coefficients, while for nonlinear relationships (B), the complexity of the curve is described by the effective degrees of freedom (edf).} 
\label{tab:gamm}
\begin{tabular}{lrrrr}
   \hline
\emph{A: Linear effects} & Coefficient & Std. Err. & t-value & p-value \\ 
\hline
% \multicolumn{5}{l}{\emph{A: Linear effects}}\\[.1cm]
  $\quad$ (Intercept) & 37.3052 & 0.4541 & 82.1536 & $<$ 0.0001 \\ 
  $\quad$ February & -7.2787 & 0.3559 & -20.4496 & $<$ 0.0001 \\ 
  $\quad$ March & -11.6052 & 0.4169 & -27.8335 & $<$ 0.0001 \\ 
  $\quad$ April & -14.6751 & 0.5173 & -28.3679 & $<$ 0.0001 \\ 
  $\quad$ May & -20.5547 & 0.6097 & -33.7147 & $<$ 0.0001 \\ 
  $\quad$ June & -22.0366 & 0.6978 & -31.5800 & $<$ 0.0001 \\ 
  $\quad$ July & -23.2506 & 0.7287 & -31.9076 & $<$ 0.0001 \\ 
  $\quad$ August & -24.5223 & 0.6813 & -35.9908 & $<$ 0.0001 \\ 
  $\quad$ September & -22.7519 & 0.5743 & -39.6143 & $<$ 0.0001 \\ 
  $\quad$ October & -18.9167 & 0.4687 & -40.3605 & $<$ 0.0001 \\ 
  $\quad$ November & -15.8692 & 0.4005 & -39.6225 & $<$ 0.0001 \\ 
  $\quad$ December & -11.4140 & 0.3541 & -32.2379 & $<$ 0.0001 \\ 
  $\quad$ Altitude & -0.0023 & 0.0012 & -1.9378 & 0.0526\\[.3cm] 
\hline
  \emph{B: Nonlinear effects}  & edf & Ref.df & F-value & p-value \\ 
\hline
% \multicolumn{5}{l}{\emph{B: nonlinear effects}}\\[.1cm]
  $\quad$ WE\_temp\_2m & 8.5851 & 8.5851 & 267.3673 & $<$ 0.0001 \\ 
  $\quad$ WE\_tot\_precipitation & 7.2891 & 7.2891 & 210.3836 & $<$ 0.0001 \\ 
  $\quad$ WE\_rh\_mean & 8.0060 & 8.0060 & 602.4805 & $<$ 0.0001 \\ 
  $\quad$ WE\_wind\_speed\_100m\_mean & 6.5219 & 6.5219 & 214.0715 & $<$ 0.0001 \\ 
  $\quad$ WE\_blh\_layer\_max & 8.8552 & 8.8552 & 278.6934 & $<$ 0.0001 \\ 
  $\quad$ LI\_pigs\_v2 & 7.9422 & 7.9422 & 7.2958 & $<$ 0.0001 \\ 
  $\quad$ LI\_bovine\_v2 & 3.7865 & 3.7865 & 1.2020 & 0.2962 \\ 
  $\quad$ LA\_hvi & 8.4121 & 8.4121 & 23.1429 & $<$ 0.0001 \\ 
  $\quad$ LA\_lvi & 7.4114 & 7.4114 & 82.1857 & $<$ 0.0001 \\ 
  $\quad$ Longitude,Latitude & 27.7758 & 32.0000 & 19.0023 & $<$ 0.0001 \\ 
   \hline
\end{tabular}
\end{table}

\subsubsection{GAMM}

\begin{figure}
     \centering
         \centering
         \includegraphics[width=1\textwidth]{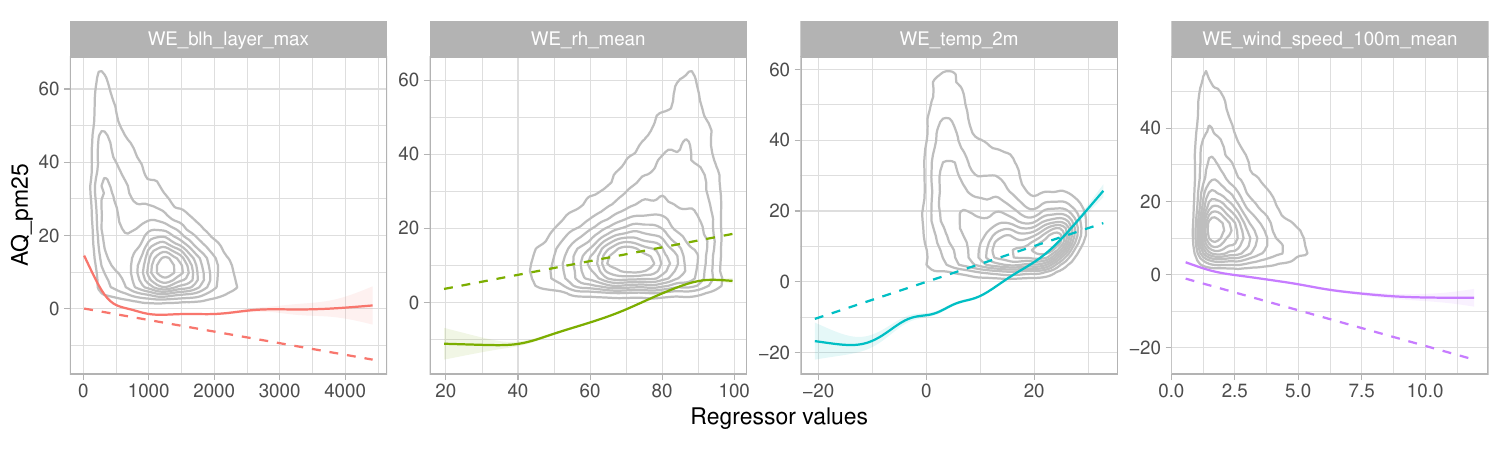}
         \caption{Smoothing splines (continuous lines), including their 95\% confidence intervals (coloured shadows), and HDGM regression coefficients (dotted lines) for relevant weather regressors (i.e., boundary layer height, relative humidity, temperature, and wind speed). Grey contour lines represent the estimated two-dimensional kernel densities of the PM$_{2.5}$ concentrations and the corresponding regressors.}
         \label{fig:spline_r}
\end{figure}

The estimated coefficients of our second model, the GAMM, are presented in Table \ref{tab:gamm}. In the first section of the table, the estimated coefficients for the linear part of the model (including the monthly fixed effects) are reported, while the second part summarises the effective degrees of freedom of the nonlinear effects as a measure of complexity/non-linearity. Compared to HDGM, the monthly fixed effects are slightly smaller, indicating that the seasonal variation is better captured by the weather variables, which enter the model nonlinearly. Complex nonlinear relationships with large degrees of freedom characterised the smooth terms of the cubic regression splines. All of them are significant except for the density of bovine. To illustrate the difference between the linear effects in the HDGM and the nonlinear effects in GAMM, we depict some selected regression splines in Figure \ref{fig:spline_r} along with the estimated linear functions of the HDGM. These curves correspond to the marginal effect of variables neglecting the spatiotemporal correlation (i.e., without the influence of neighbouring sites). In general, we observe a similar tendency for both models. An exceptional notice would be the height of the boundary layer, which has a negative effect for up to 500 kilometres, and afterwards, the effect changes to be positive. By contrast, the effect is negative for the HDGM, which mimics the effect in the areas where most observations are located. The grey contour lines in Fig. \ref{fig:spline_r} additionally illustrate the estimated kernel density of the couple (PM$_{2.5}$, WE\_regressor). We note that the confidence intervals around the fitted curves are smaller in areas with higher density.

\begin{figure}
     \centering
         \includegraphics[width=0.9\textwidth]{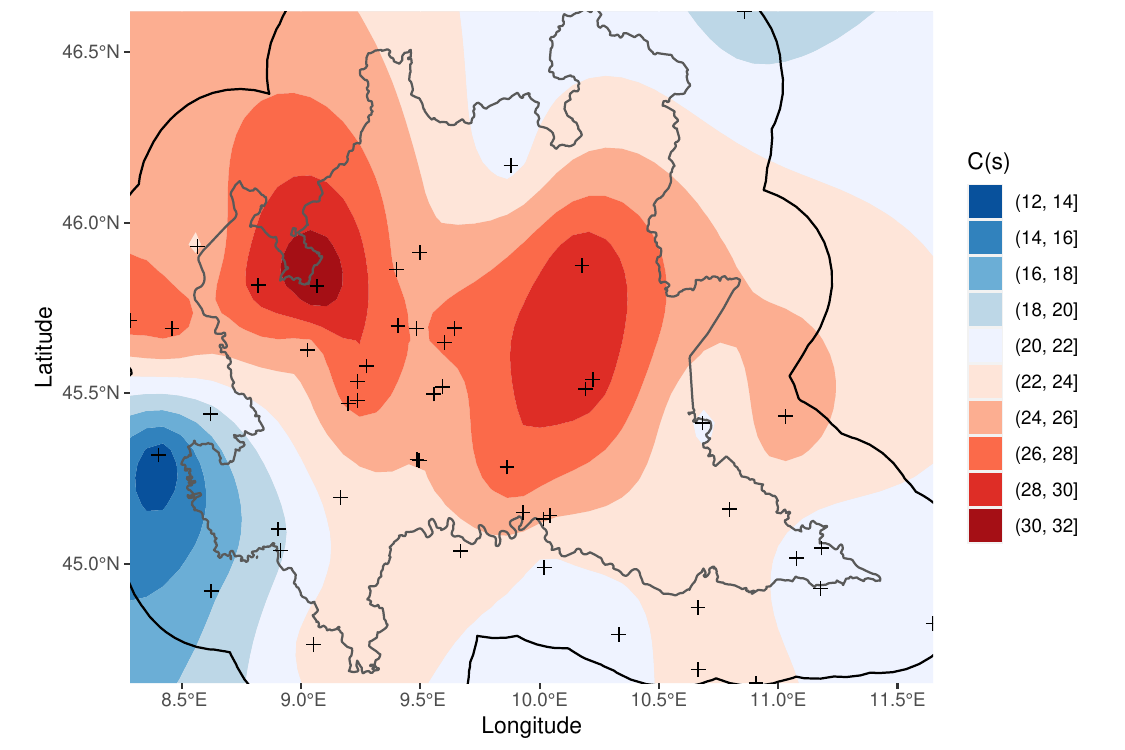}
         \caption{Estimated smoothing spline $\hat{C}(s)$ of GAMM, which corresponds to the PM$_{2.5}$ predicted on a regular grid using the large scale of GAMM, where all regressors are set to 0. Stations are marked with a black cross, Lombardy boundaries are shown in grey, and the black line marks the surrounding buffer zone.}
         \label{fig:splines_s}
\end{figure}

The GAMM smooth spatial surface $C(s)$ is displayed in Fig. \ref{fig:splines_s}. This smoothing spline $C(s)$ capturing the spatial dependence identifies correlated areas. Our study shows higher concentrations of PM$_{2.5}$ in the area of Como and the area of Brescia, while in the southwest, corresponding to the Ligurian border, lower concentrations. Furthermore, the estimated range parameter of the exponential covariance function is $\hat{\theta}_{GAMM} = 1.16^{\circ}$, which corresponds to approximately 110 km. Hence, it is in a similar range to the other two models. Furthermore, the autoregressive parameter is estimated as $\hat{g}_{GAMM} = 0.67$, similar to HDGM and indicating a medium temporal persistence across one day. In this sense, the models show similar spatiotemporal dynamics as the HDGM.

% GAMM small-scale component is represented by a first-order autoregressive process with coefficient $g_{GAMM}$ for temporal effects and a smooth spatial surface $C(s)$ for the spatial effects. The estimated function $C(s)$ is displayed in Fig. \ref{fig:splines_s}. The smoothing spline $C(s)$ capturing the spatial dependence identifies correlated areas where predictions are over (or under) estimated. In our study, the spatial dependence determines in the area of Como and the area of Brescia higher concentrations of PM$_{2.5}$, while in the southwest, corresponding to the Ligurian border, lower concentrations. Furthermore, the estimated decay parameter of the exponential covariance function is $\hat{\theta}_{GAMM} = 1.16^{\circ}$, which corresponds to approximately 110 km. Hence, it is in a similar range to the other two models. Furthermore, the autoregressive parameter is estimated as $\hat{g}_{GAMM} = 0.67$, indicating a medium temporal persistence across one day. In this sense, the models show similar spatiotemporal dynamics as the HDGM. 

\subsubsection{RFSTK}

\begin{figure}
    \centering
    \includegraphics[width=.5\textwidth]{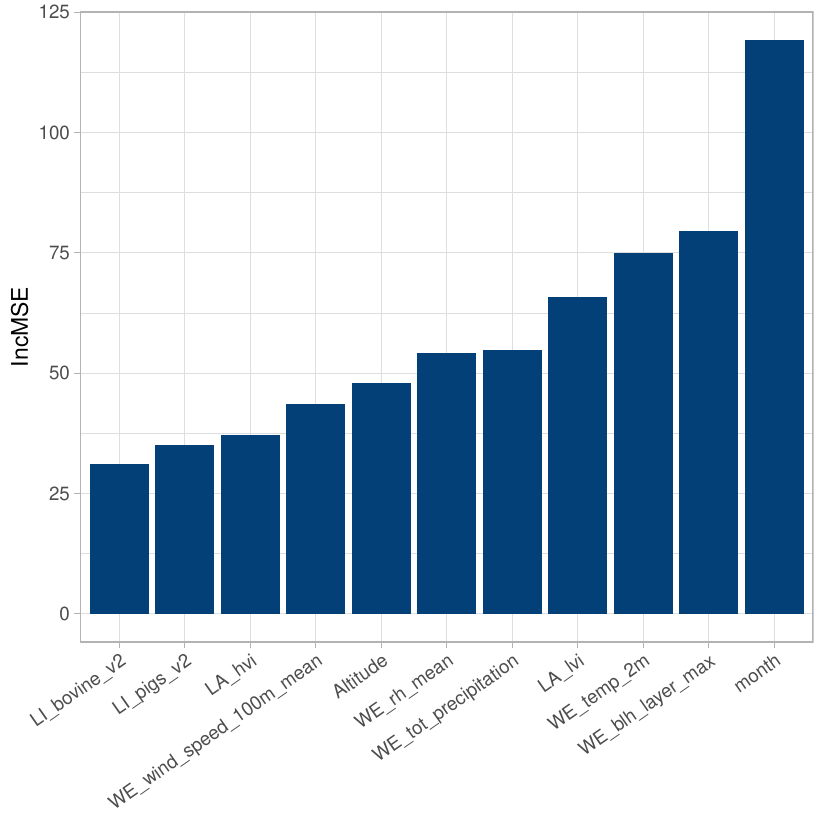}
    \caption{Variable Importance Factor (measured as IncMSE) for the 11 selected features in the large-scale component of the RFSTK.}
    \label{fig:vif1}
\end{figure}

The interpretability of the RFSTK model can be challenging due to its intricate nature. However, the variable importance factor (VIF) can help to identify the most important variables. To determine the VIF, the mean decrease accuracy technique is employed, which assesses the variables' importance by measuring the increase in MSE (IncMSE) when the values of the regressors are permuted \citep{breiman2001random}. The results are depicted in Fig. \ref{fig:vif1}.

Like the previous models, the random forest cannot fully capture the seasonality by the included (weather) regressors, as we can see by the high importance of the monthly effects. The most important variables after the month are the height of the boundary layer, the temperature and the (low) vegetation index. The bovine density is the least important variable, and indeed, it was not significantly different from zero for the other two models.

For the RFSTK, the spatiotemporal dependence is captured by considering a spatiotemporal Gaussian process with a separable exponential covariance function for space and time. The corresponding variogram model is fitted to the RF residuals. For our analysis, we obtained exponential decay parameters of $\hat{\theta}_{RFSTK_t} = 0.78$ days and $\hat{\theta}_{RFSTK_s} = 0.48^{\circ}$ for time and space, respectively, the latter corresponds to approximately 50 km. The spatial range parameters are smaller than HDGM because the large scale (RF) explains more variation than the linear large-scale term of HDGM, as shown in Tab. \ref{tab:residuals}.

\subsection{Model comparison} 

Intriguing observations arise when comparing the results of the three models. First, the temporal variation plays an essential role in all three models, as evidenced by the magnitude of the temporal dummy coefficients in HDGM and GAMM (with January as the worst month for air quality) and `month' receiving the highest ranking in the VIF of RFSTK. This suggests that weather variables alone are insufficient for capturing all seasonal variability, even considering the most important weather variables. Interestingly, temperature, which is usually negatively associated with PM$_{2.5}$, was found to be positively related to PM$_{2.5}$ conditional on the month. This is due to the opposing effect of temperature and monthly indicator variables. Furthermore, surprisingly, the significance tests in both HDGM and GAMM on livestock densities suggested that bovine livestock density is not related to PM$_{2.5}$ concentrations, while pig density is.

\begin{figure}
    \centering
    \includegraphics[width=1\textwidth]{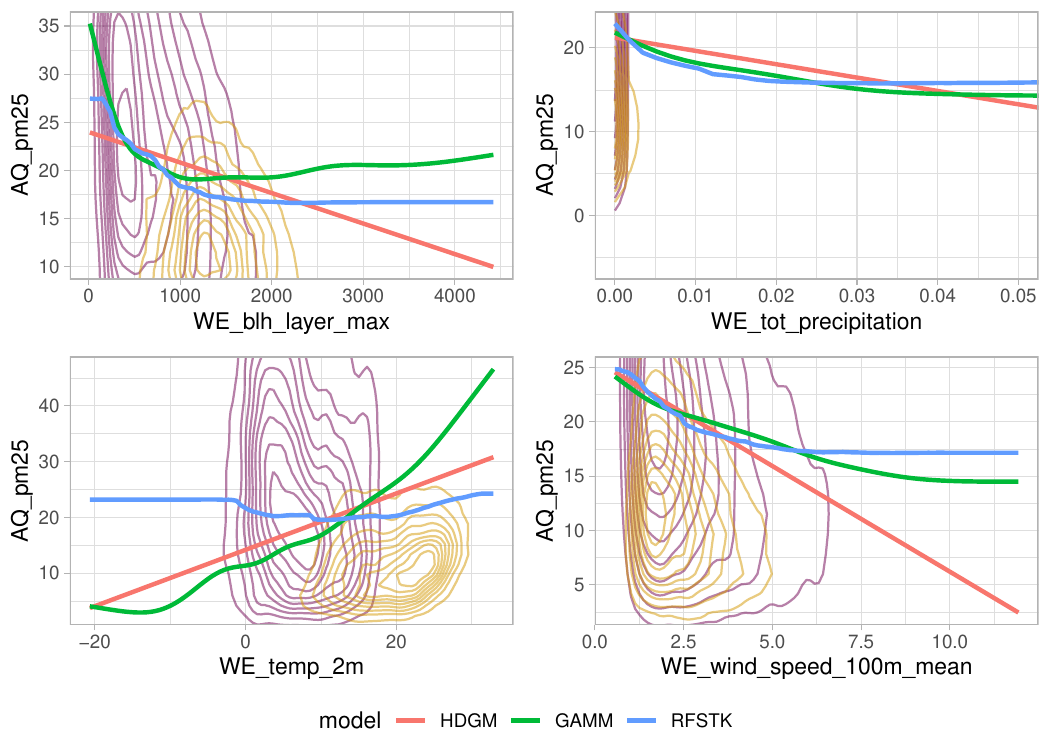}
    \caption{PDP calculated on the large-scale of all three models. Contour lines represent kernel density estimation of the couple (PM$_{2.5}$, WE\_regressor), in yellow for hot months and purple for cold months.}
    \label{fig:comparison2}
\end{figure}

In Figure \ref{fig:comparison2}, we depict partial dependence plots (PDPs) \citep{friedman2001greedy} for selected covariates of all three models along with two-dimensional kernel density estimates of the pairs of PM$_{2.5}$ concentrations and the corresponding regressors in the hot (summer and spring) and cold (autumn and winter) months as yellow and purple contours, respectively. PDPs allow us to compare the relationships identified within the large-scale of each model and highlight the transition from linear to more complicated relationships. Contrary to the marginal effects, these PDPs account for the typical range of the other predictors. This is accomplished by associating a fixed value of a regressor across all observations with the mean of predicted PM$_{2.5}$ concentrations. The mean of prediction is calculated for different fixed values of the regressor, typically moving from the minimum to the maximum on an equidistant grid. In other words, PDPs show how the predicted outcome of the changes as a single predictor variable is varied while all other variables are held constant. 
%They can be a helpful tool for understanding the relationship between a particular predictor variable and the predicted response variable. 

% How the results of RF can be used for model selection of interpretable models (in a fair way) (23)

% \textcolor{red}{\textbf{In the following, GAMM seems not mentioned, even if it represents the 'transition' traditional model to have nonlinearities and flexibility in the large-scale part.} }\\

It is worth noting that all three models demonstrate similar trends, even though they have different levels of flexibility. For example, they all show that PM$_{2.5}$ concentrations are higher in cold periods. However, slight differences exist in the models' behaviour, especially for temperature. This suggests there may be nonlinear influences or interactions between temperature and other variables. For example, temperature may have a different impact on PM$_{2.5}$ concentrations in different altitudes or seasons. RFSTK can capture these interactions more effectively than the other two models, which is why its PDP is flatter. This suggests that RFSTK is better at capturing the complex relationships between PM$_{2.5}$ concentrations and other variables.

This finding has important implications for the development of air quality models. It suggests that machine learning (ML) techniques can be used to improve the performance and interpretability of classic geostatistical approaches, such as HDGM or GAMM. This is because ML techniques can identify nonlinearities and interactions that are difficult to identify using traditional methods. In the second step, the more interpretable models could include the nonlinear effects and interactions, e.g. GAMM. The comparison of PDPs also highlights the complementary nature of ML techniques and classic approaches. ML techniques are better at capturing complex relationships, while traditional approaches are easier to interpret and allow for straightforward uncertainty estimation. Thus, we advertise combining ML techniques and classic approaches for modelling and predicting PM concentrations.

\section{Conclusion}\label{sec::conclusion}

This study compares three statistical models to model and predict Lombardy's daily PM$_{2.5}$ concentrations and simultaneously provide an intuitive interpretation of the influencing factors. The models considered are HDGM, GAMM and RFSTK. All three models used are designed to handle spatiotemporal data, although each employs different methods to model external factors and spatiotemporal dependence. The models can generally be divided into large-scale components, small-scale spatiotemporal effects, and measurement and modelling errors. The large-scale components account for external influences, whereas the small-scale components model the spatiotemporal correlation, and the modelling errors contain the unexplained variation of the process.

The large-scale component of the three models showed significant monthly fixed effects for all three models, with negative coefficients for all months. As expected, January was confirmed to be the month with the worst air quality. Furthermore, we used partial dependence plots to compare relationships within the large-scale of the three models and highlight the transition from linear to more complex relationships. The generalised additive mixed model and the random forest approach exhibit similar patterns as they can handle nonlinear relationships. The geostatistical model is constrained by its linear specification, which fits in areas with sufficiently many observations of the covariates. At the same time, linear specification prevents making unreliable predictions -- even in regions with few observations for the model estimation. Thus, the hidden dynamic geostatistical showed the best performances in the cross-validation study with an average RMSE of 6.31 $\mu g / m^{3}$.

By comparing marginal effects, it is possible to better understand potential interactions and nonlinearities, thereby improving the model's specification. Indeed, the HDGM can produce good results due to its ability to incorporate latent variables. Still, at the same level of predictive performances, it is preferable to have a model that explains more on a large scale, extending the degree of interpretability. Thus, it can be highly beneficial to detect nonlinear behaviours or interactions by machine learning, which can then be used to improve the specification of the ``simpler'' but faster model integrated with a more efficient spatiotemporal correlation structure. ML techniques and classic statistical models can be used in complementary ways.

The comparison of models in the field of air quality highlighted that the spatiotemporal correlation is a crucial aspect that requires careful consideration. However, this correlation is also very sensitive. If not handled properly, it can lead to overfitting the model to the specific data used, thereby hindering its ability to generalise the discovered relations. This limitation was illustrated by the discrepancy in the performance of the HDGM model when evaluated on the entire data set versus when assessed using the leave-one-station-out cross-validation approach.

% Our comparative analysis highlighted the differences in the performance of classic geostatistical methods and machine learning techniques such as random forests. We considered the spatiotemporal dependence in all models. 
In conclusion, our findings suggest that classic approaches, such as the hidden dynamic geostatistical model, yield the best predictive performance while being computationally efficient. However, more complex algorithms like random forest can enhance the identification of nonlinear and interaction effects. Therefore, these methods can be used complementary, ushering in a new era where newly developed techniques support traditional and well-established practices.

\section{Acknowledgement}

This research was funded by Fondazione Cariplo under the grant 2020–4066 ``AgrImOnIA: the impact of agriculture on air quality and the COVID-19 pandemic'' from the ``Data Science for science and society'' program.

% \bibliography{Bib1}% common bib file

\end{document}